# Surface Plasmon Enhanced Photoconductance of Gold Nanoparticle Arrays with Incorporated Alkane Linkers

M. A. Mangold[1], C. Weiss[1], M. Calame[2] and A. W. Holleitner[1],*

1) Walter Schottky Institut and Physik Department, Technische Universität München, Am Coulombwall 3, D-85748 Garching, Germany.

2) Departement für Physik, Universität Basel, Klingelbergstrasse 82, CH-4056 Basel, Switzerland.

We report on a photoconductive gain effect in two-dimensional arrays of gold nanoparticles, in which alkane molecules are inserted. The nanoparticle arrays are formed by a self-assembly process from alkanethiol-coated gold nanoparticles, and subsequently they are patterned on a Si/SiO$_2$ chip by a microcontact printing technique. We find that the photoconductance of the arrays is strongly enhanced at the frequency of the surface plasmon of the nanoparticles. We interpret the observation as a bolometric enhancement of the conductance of the nanoparticle arrays upon excitation of the surface plasmon resonance.

The excitation of surface plasmons in two-dimensional arrays of metal nanoparticles can lead to a strong enhancement of the electromagnetic field at the surface of the particles at visible and near-infrared wavelengths.[1] Recent experimental and theoretical work on the plasmonic enhancement effect focused on surface enhanced Raman spectroscopy (SERS)[2-4] and bio-sensing applications[5,6] as well as the photonic band gap[7] and the photoelectrochemical properties of such nanoparticle arrays.[8] In the context of molecular electronics, densely packed two-dimensional nanoparticle arrays have been exploited to electrically contact molecules.[9,10] Optical studies revealed a strong plasmonic absorption in such nanoparticle arrays, where alkanes or conjugated molecules have been inserted.[11] Experiments with photochromic molecules, which were immobilized in the nanoparticle-arrays, showed that the optical excitation of organic photoswitches in nanoparticle arrays is possible, and that changes in the conductance of the nanoparticle arrays due to optical switching of the molecules can be observed.[12,13] However, an open question remains whether surface plasmon resonances can be exploited to induce charge transport across nanoscale metal junctions with molecules embedded.[14]

Here, we experimentally investigate the photoconductance properties of well ordered two-dimensional gold nanoparticle arrays consisting of millions of metal junctions, in which alkane molecules are incorporated. We measure the photoconductance of the arrays as a function of laser intensity, modulation frequency, bias voltage, spatial coordinates, and wavelength.[15,16] We find a clear enhancement of the photoconductance at the surface plasmon resonance of the nanoparticle arrays. We interpret the findings such that the conductance of the arrays is bolometrically increased when the surface plasmons in the nanoparticle arrays are optically excited. We find good agreement between the data and a simple model, which estimates the conductance change of the nanoparticle arrays due to an optically induced temperature increase.

Starting point for the two-dimensional nanoparticle arrays is the synthesis of gold nanoparticles with a diameter of approximately 9.2 nm following ref. [17]. The nanoparticles are then covered with a layer of alkane molecules by incubation with octanethiols dissolved in ethanol for 24 hours (all chemicals are purchased from Sigma-Aldrich and used as delivered). The alkane coated nanoparticles self-assemble into two-dimensional hexagonal arrays at an air-water interface.[18] Subsequently, 25 μm wide stripes of a nanoparticle array are patterned on a $SiO_2$ chip via a microcontact printing technique.[19] In a shadow mask evaporation step, these stripes are contacted by macroscopic gold electrodes with a separation of ~8 μm, leaving a contacted area with dimensions of $25 \times 8$ μm$^2$. The electrodes are electrically connected to a chip carrier with gold wires using a wedge bonder. The scanning electron microscope (SEM) image in Fig. 1(a) shows a high resolution image of such a nanoparticle array. Predominantly, the nanoparticles are arranged in a hexagonal array with a lattice constant of approximately 12 nm and a particle-particle distance of about 2 – 3 nm.[9] Fig. 1(b) shows a typical room temperature current-voltage characteristic of the nanoparticle arrays, which is measured in a two-terminal configuration. The corresponding array exhibits an almost ohmic behavior in the voltage range from -10 to +10 V with a conductance of 5.1 nS.

All photoconductance measurements are carried out at room temperature in a vacuum chamber at a pressure of about $10^{-5}$ mbar. Optical excitation occurs by focusing the light of a mode-locked titanium:sapphire laser with a repetition rate of 76 MHz through the objective of a microscope onto the nanoparticle array. The laser can be tuned in the wavelength range 700 nm < $\lambda_{PHOTON}$ < 1000 nm. Excitation of a non-linear optical fiber produces a broad spectrum of light, from which the wavelengths between 550 nm and 700 nm can be selected with filters. Second harmonic generation in a $\beta$-BaB$_2$O$_4$ (BBO) crystal yields the wavelengths between 400 and 500 nm. With a spot diameter of ~2 µm the light intensity $I_{OPT}$ is on the order of 1 kW/cm$^2$ for all wavelengths. For the photoconductance measurements, we chop the laser at a frequency $f_{CHOP}$. The resulting current $I_{PHOTO}(\lambda_{PHOTON}, f_{CHOP}) = I_{ON}(\lambda_{PHOTON}, f_{CHOP}) - I_{OFF}$ across the sample with the laser being "ON" or "OFF", respectively, is amplified by a current-voltage converter and detected with a lock-in amplifier utilizing the reference signal provided by the chopper. Fig. 1(c) depicts the optically induced current $I_{PHOTO}$ measured on a nanoparticle array as a function of time, while the laser light is switched on and off. We infer from this measurement that the photoconductance does not change over time even for an illumination period of several minutes. Fig. 1(d) shows $I_{PHOTO}$ as a function of $V_{SD}$. We find that $I_{PHOTO}$ depends linearly on $V_{SD}$ and we detect no finite value of $I_{PHOTO}$ at zero bias. Thus, the measured signal is due to an optically induced change of conductance, and we use the term "photoconductance", i.e., $G_{PHOTO} = I_{PHOTO}/V_{SD}$, to describe the observed phenomenon.

In Fig. 2(a), the spatially resolved photoconductance of the area shown in Fig. 2(b) is depicted. To this end, $G_{PHOTO}$ is measured, while the position of the sample is moved in the x-y-plane in steps of 2 µm. We observe a photoconductance signal for an area of exactly the size of the electrically contacted stripes of the nanoparticle array. The measured photoconductance ranges from $G_{PHOTO}$ ~5 fS, when illuminating the source and drain electrodes, to a value of $G_{PHOTO}$ ~40 fS on the area covered by the nanoparticle array. In Fig. 3(a) we show the photoconductance for different excitation wavelengths. We observe a clear maximum of the photoconductance at approximately 600 nm [triangle in Fig. 3(a)], which coincides with the surface plasmon resonance of two-dimensional arrays made out of octane coated nanoparticles.[11] Therefore, the strong maximum of the photoconductance at that wavelength suggests that surface plasmons play an important role in the creation of the photoconductance.

In order to describe our observations, we start with the Drude-Lorentz-Sommerfeld model where the dielectric function of an isolated Au nanoparticle is given by $\varepsilon_{NP} = \varepsilon_{1,NP} + i\varepsilon_{2,NP}$, with $\varepsilon_{1,NP} = 1 - \frac{\omega_p^2}{\omega^2 + \Gamma^2} + \varepsilon_{1,core}$ and $\varepsilon_{2,NP} = 1 - \frac{\omega_p^2 \Gamma}{\omega(\omega^2 + \Gamma^2)} + \varepsilon_{2,core}$ the real and imaginary parts. Here, $\omega_p$ denotes the bulk plasmon frequency of gold, and $\varepsilon_{1,core}$ ($\varepsilon_{2,core}$) is the real (imaginary) part of the core electron contribution to the dielectric response of the nanoparticles. $\Gamma(R_{NP})$ is a size-dependent damping constant, which is given by $\Gamma(R_{NP}) = \frac{v_F}{l_\infty} + P v_F / R_{NP}$, with $v_F$ the Fermi velocity in gold, $l_\infty$ the mean free path of conduction electrons in gold, $R_{NP}$ the nanoparticle diameter, and $P$ a proportionality factor.[20] The absorption of a two-dimensional layer of nanoparticles in a dielectric medium can be calculated as the absorption $A_{eff}$ of an effective medium and reads as $A_{eff} = 1 - e^{-\kappa d_{NP}}$ with $d_{NP} = 12$ nm the thickness of a layer of alkane-coated nanoparticles and the absorption coefficient[21]

$$\kappa = \omega \text{Im}[\varepsilon_{eff}]/c n_{eff}, \tag{1}$$

where $\hbar\omega$ is the energy of the absorbed photons and $c$ is the velocity of light in vacuum. $\varepsilon_{\text{eff}}$ and $n_{\text{eff}} = \text{Re}[\sqrt{\varepsilon_{\text{eff}}}]$ are the dielectric function and the refractive index of the effective medium. The Maxwell-Garnett effective medium theory predicts[21-23]

$$\varepsilon_{\text{eff}}(\omega) = \varepsilon_{\text{m}} \frac{\varepsilon_{\text{NP}}(\omega)(1+2f)+2\varepsilon_{\text{m}}(1-f)}{\varepsilon_{\text{NP}}(\omega)(1-f)+\varepsilon_{\text{m}}(2+f)} \tag{2}$$

with $\varepsilon_{\text{m}}$ the dielectric constant of the medium surrounding the nanoparticles and $f$ the volume fraction of the gold. Following ref. [11], we use $f = 0.34$ for arrays of C8 coated nanoparticles, $\varepsilon_{\text{m}} = 2.5$ for the alkanes bound to the nanoparticles, and $P = 2.0$. In addition, we use $\omega_{\text{p}} = 1.37*10^{16}$ Hz,[24] $v_{\text{F}} = 1.4*10^{6}$ m/s,[25] $l_\infty = 29.4$ nm,[25] and take the values of $\varepsilon_{1/2,\text{core}}$ according to Ref. [26]. As a result, we obtain an absorption spectrum with a maximum absorption at a wavelength of 595 nm [triangle in Fig. 3(b)], which is consistent with recent measurements.[11] Hereby, the calculation corroborates the interpretation that the photoconductance maximum at 600 nm results from a surface plasmon resonance in the two-dimensional gold nanoparticle arrays.

In the following, we demonstrate that the photoconductance maximum at the plasmon resonance is consistent with a bolometrically induced conductance. To this end, we show in Fig. 3(c) that an increase of the bath temperature $T_{\text{BATH}}$ gives rise to an enhanced (dark) sheet conductance $G_\blacksquare = \frac{dI_{\text{SD}}}{dV}\frac{l}{w}$ with $l$ the length and $w$ the width of the array. We find an exponential dependence of $G_\blacksquare$ on the inverse temperature, as it is expected for a nanoparticle array at temperatures above the Coulomb blockade regime.[27] From a linear approximation of the measurement at 300 K, we deduce the gradient of the sheet conductance to be $dG_\blacksquare/dT_{\text{BATH}} = +2.4$ pS/K. Inspired by a study by Govorov et al., which demonstrates that the heat generation in nanoparticles is strongly enhanced by excitation of the surface plasmon resonance, we calculate the temperature increase of the nanoparticle array due to the absorbed light.[28] To that end, we assume that thermal equilibrium is reached on time scales much shorter than the typical illumination times in our experiment, which will be justified below. We equate the absorbed light intensity $I_{abs} = A_{eff}I_{opt}$ with the heat conducted away from the array through the underlying $SiO_2$ layer. Since the heat conductivity of $SiO_2$ is lower by more than a factor 100 compared to the heat conductivity of Si, we do not consider the heat conductance in the bulk of the silicon chip.[26] Also, we neglect the heat conductance inside the array due to its small cross section. The heat conductance per unit area of $SiO_2$ is given by $Q = \lambda \frac{\Delta T}{d_{SiO_2}}$, with $\lambda$ the thermal conductivity of $SiO_2$, $d_{SiO_2}$ the thickness of the $SiO_2$ layer, and $\Delta T$ the temperature increase in the nanoparticle array.[29] We evaluate the temperature increase of the nanoparticle array at the surface plasmon resonance to be $\Delta T$ = 0.55 K (for $I_{opt}$ = 1.3 kW/cm², $A_{eff}$ = 0.31 at 600 nm, $\lambda$ = 1.1W/(m K) for $SiO_2$,[30] and $d_{SiO_2}$ = 150 nm). We then describe the nanoparticle array by 4 times 12.5 resistors, each of which has an area of 2x2 µm², and they are connected in a rectangular lattice geometry. In the experiment, only one resistor at a time is excited by the laser beam, and in turn, its conductance is increased by $dG_\blacksquare/dT_{BATH} \times \Delta T$ = +2.4 pS/K × 0.55 K = 1.32 pS. We then calculate the conductance of the whole network for one resistor being illuminated, and subtract the dark conductance to determine the photoconductance value at 600 nm to be $G_{PHOTO}$ = 81 fS. The latter value is in the same order of magnitude as the experimentally determined photoconductance signal of 200 fS at 600 nm [Fig. 3(a)]. The discrepancy between both values can be explained by the fact, that the simple model does not consider the rather bad thermal coupling between the nanoparticles and the SiO$_2$ substrate. Hereby, a further heat accumulation is very plausible.

In Fig. 3(d), we show the photoconductance as a function of the chopper frequency $f_{CHOP}$. We do not find a dependence of the photoconductance on $f_{CHOP}$ for frequencies between 300 and 4000 Hz. This reveals that the effect causing the photoconductance occurs on a time-scale shorter than ~ (4000 Hz)$^{-1}$ = 250 µs. In Ref. [28] it is reported that the heat accumulation in metal nanoparticle layers builds up faster than 1 µs. From the complete lack of chopper frequency dependence of the photoconductance, we conclude that the system reaches thermal equilibrium faster than the experimentally accessible time of 250 µs. Finally, we note that the observed photoconductance signal depends linearly on $I_{OPT}$ over two orders of magnitudes (data not shown). This observation is consistent with the above interpretation that the photoconductance is caused by a bolometrically enhanced conductance of the nanoparticle arrays.

In summary, we report on the photoconductance properties of two-dimensional arrays of gold nanoparticles, which are formed by a combination of a self-assembly process and a microcontact printing technique. The photoconductance of the nanoparticle arrays is strongly enhanced through the excitation of surface plasmons. We interpret the observation by a bolometric enhancement of the conductance of the nanoparticle array.

We thank C. Schönenberger for fruitful discussions, and we gratefully acknowledge support from the German excellence initiative via the "Nanosystems

Initiative Munich" (NIM) and the "Center for NanoScience" (CeNS). MC acknowledges the National Center of Competence in Research 'Nanoscale Science' and the GEBERT RUF STIFTUNG.

**Fig. 1** (color online): (a) Scanning electron microscope (SEM) image of an array from gold nanoparticles coated with octane thiols. (b) Current-voltage characteristic of a nanoparticle array at room temperature. (c) Optically induced current as a function of laboratory time ($V_{SD}$ = 10 V, $f_{CHOP}$ = 814 Hz, $I_{opt}$ = 0.4 kW/cm$^2$, $\lambda_{PHOTON}$ = 620 nm). The (red) highlighted areas indicate times where the laser light is on. (d) Light induced current as a function of source-drain voltage ($f_{CHOP}$ = 1716 Hz, $I_{opt}$ = 4.1 kW/cm$^2$, $\lambda_{PHOTON}$ = 620 nm).

**Fig. 2** (color online): (a) Photoconductance as a function of the laser spot position ($V_{SD}$ = 10 V, $f_{CHOP}$ = 812 Hz, $I_{opt}$ = 2 kW/cm$^2$, $\lambda_{PHOTON}$ = 562 nm). Source and drain electrode are drawn schematically as a guide to the eye. (b) SEM-graph of a 25 μm wide stripe of nanoparticle array, contacted by gold electrodes with a distance of 8 μm.

**Fig. 3** (color online): (a) Photoconductance for different excitation wavelengths ($V_{SD}$ = 10 V, $f_{CHOP}$ = 812 Hz, $I_{OPT}$ = 1.3 kW/cm$^2$). (b) Calculated absorption spectrum of the nanoparticle array according to Eqs. (1) and (2). (c) Logarithmic plot of the sheet conductance of the nanoparticle array without laser excitation vs. the inverse bath temperature at $V_{SD}$ = 10V. (d) Photoconductance as a function of the chopper frequency ($V_{SD}$ = 10 V, $I_{opt}$ = 0.4 kW/cm$^2$, $\lambda_{PHOTON}$ = 620 nm).

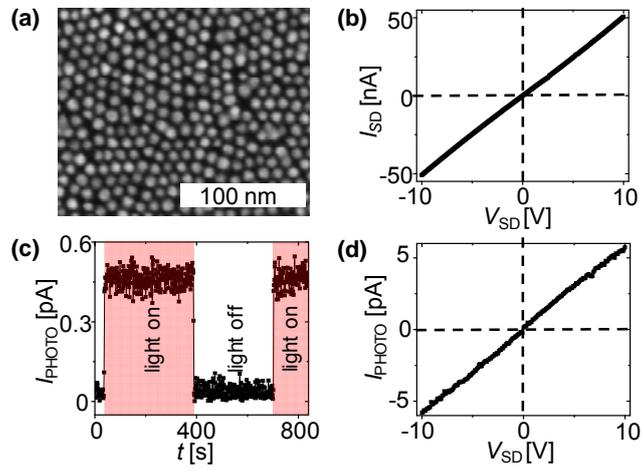

Figure 1

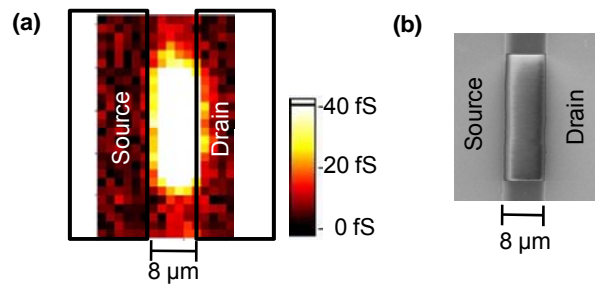

Figure 2

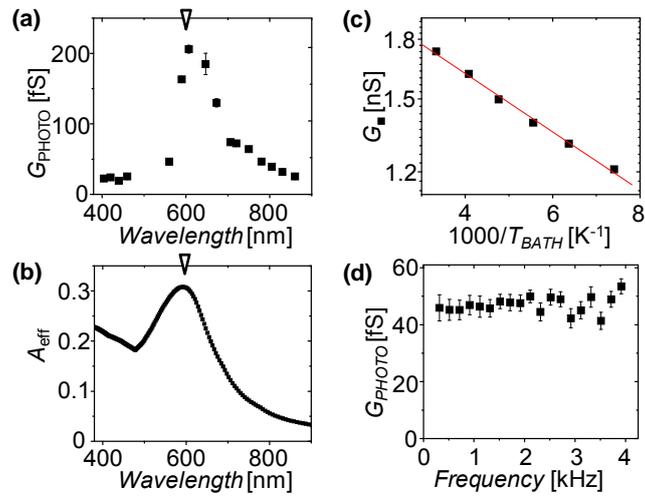

Figure 3